\documentclass{iau}
\usepackage{graphicx}

\title[IAUS292~~Probing the cool ISM in galaxies] 
{Probing the cool ISM in galaxies via 21\,cm \mbox{H\,{\sc i}} absorption}

\author[J.~R. Allison, E.~M. Sadler, S.~J. Curran \& S.~N. Reeves]   
{J.~R. Allison$^{1,\ast}$,
E.~M. Sadler$^{1,2}$, 
S.~J. Curran$^{1,2}$
\and S.~N. Reeves$^{1,2,3}$}

\affiliation{$^1$Sydney Institute for Astronomy, School of Physics
  A28, \\University of Sydney, NSW 2006, Australia \\[\affilskip]
  $^{2}$ARC Centre of Excellence for All-sky Astrophysics (CAASTRO)
  \\[\affilskip]$^{3}$CSIRO Astronomy \& Space Science, P.O. Box 76,
  Epping NSW 1710, Australia\\[\affilskip]$^{\ast}$email: {\tt
    jra@physics.usyd.edu.au}}

\pubyear{2012}
\volume{292}  
\pagerange{xxx--xx}
\jname{Molecular Gas, Dust, and Star Formation in Galaxies}
\editors{T. Wong \& J. Ott, eds.}
\begin{document}

\maketitle

\begin{abstract}
  Recent targeted studies of associated \mbox{H\,{\sc i}} absorption
  in radio galaxies are starting to map out the location, and
  potential cosmological evolution, of the cold gas in the host
  galaxies of Active Galactic Nuclei (AGN). The observed 21\,cm
  absorption profiles often show two distinct spectral-line
  components: narrow, deep lines arising from cold gas in the extended
  disc of the galaxy, and broad, shallow lines from cold gas close to
  the AGN (e.g. Morganti et al. 2011). Here, we present results from a
  targeted search for associated \mbox{H\,{\sc i}} absorption in the
  youngest and most recently-triggered radio AGN in the local universe
  (Allison et al. 2012b). So far, by using the recently commissioned
  Australia Telescope Compact Array Broadband Backend (CABB; Wilson et
  al. 2011), we have detected two new absorbers and one
  previously-known system. While two of these show both a broad,
  shallow component and a narrow, deep component (see
  Fig.\,\ref{fig1}), one of the new detections has only a single
  broad, shallow component. Interestingly, the host galaxies of the
  first two detections are classified as gas-rich spirals, while the
  latter is an early-type galaxy. These detections were obtained using
  a spectral-line finding method, based on Bayesian inference,
  developed for future large-scale absorption surveys (Allison et
  al. 2012a).

  \keywords{methods: data analysis, galaxies: active, radio lines:
    galaxies}
\end{abstract}



\begin{figure}[h]
  \begin{center}
    \includegraphics[width=0.975\textwidth]{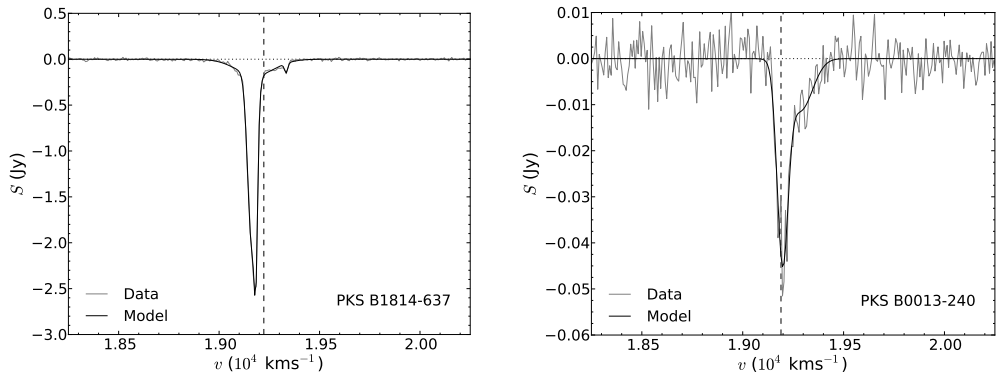} 
    \caption{21\,cm absorption in the late-type host galaxies of two
      compact radio sources.}
    \label{fig1}
  \end{center}
\end{figure}

\end{document}